# A computational intuition pump to examine group creativity: building on the ideas of others

**Ricardo Sosa**, Design and Creative Technologies, Auckland University of Technology, rsosa@aut.ac.nz

**Andy M. Connor**, Design and Creative Technologies, Auckland University of Technology, aconnor@aut.ac.nz

**Abstract**
This paper presents a computational approach to modelling group creativity. It presents an analysis of two studies of group creativity selected from different research cultures and identifies a common theme ("idea build-up") that is then used in the formalisation of an agent-based model used to support reasoning about the complex dynamics of building on the ideas of others.

*Agent-based simulation; creative teams; research methods*

Understanding and managing the group dynamics that lead to creative collaborations and creative teamwork is an important part of developing innovation capability (Francis & Bessant, 2005). Of particular interest are multidisciplinary and multicultural settings (Fruchter, 2001). Several problems are associated with methodically studying a process such as creativity, elusive by definition (Cardoso de Sousa, 2007). This paper applies agent-based simulation (ABS) as a way to model multi-level principles related to creativity and innovation (Watts & Gilbert, 2014). It illustrates how computational models can be relevant for mixed-method studies (Creswell, 2009) by modelling qualitative and quantitative aspects of group creativity, in particular the critical process of building upon the ideas of others.

**Introduction**
Aiming to develop a deeper and more structured reasoning about principles of group creativity, this paper adopts agent-based simulation (ABS) as a way to support rigorous thinking and argumentation. ABS is applied here as an 'intuition pump', defined as a resource to aid reasoning about complex subjects by harnessing intuition (Dennett, 2014). Using ASB as an 'intuition pump' allows researchers to setup, run and analyse computational systems as a way to support intuitive reasoning about difficult problems, to assess and clarify arguments and to reveal meaningful ideas about group creativity. Rather than seeking validity, or concrete recommendations for the management of creativity, ABS is used here as a tool to "pump an intuition" and to help researchers and practitioners say "*Aha! Oh, I get it!*" (Dennett, 2014).

The work presented in this paper builds on the tradition of research on creative collaboration (Diehl & Stroebe, 1987) and creative teams (Osborn, 1963). Its main motivation is to contribute to the theoretical understanding of what characterises creative teamwork, and to the practical formulation of evidence-based strategies to form and manage creative groups (Paulus & Nijstad, 2003).

Qualitative and quantitative research cultures

In creativity research two cultures or traditions exist based on two types of methods broadly defined as qualitative and quantitative (Mahoney & Goertz, 2006). Table 1 presents the main underlying differences between these two research cultures, related to how studies are framed, how participants are selected, and how data is collected and analysed.

Table 1: Differences between Qualitative and Quantitative Research Cultures -adapted from (Mahoney & Goertz, 2006).

| Qualitative methods used in creativity research | Quantitative methods used in creativity research |
| --- | --- |
| Within-case analysis | Cross-case analysis |
| Exemplary creative subjects selected | Randomised subjects selected; sampling |
| Descriptive analysis | Statistical analysis |
| Causes of observations | Effects of independent variables |
| Semantic treatment of concepts (definitions) | Numerical treatment of concepts (measurement) |
| Tends to asymmetric explanations | Tends to symmetric explanations |
| Deep data, often captures long processes | Wide scope data, often constrained to short processes |
| Personal experience is critical; tolerance for implicitness | Seeks thorough explicitness |

This paper aims to connect these two research cultures by building computational simulations to support reasoning about creativity and innovation (Watts & Gilbert, 2014). Both quantitative and qualitative lenses are used here, focusing specifically on the process of ("idea build-up", inspired by studies that represent the two research cultures and which coincide in giving high importance to the building of new ideas upon the ideas of others.

**Background**

In recent decades a rich body of knowledge has been advanced related to the formation, management and performance of creative groups (Diehl & Stroebe, 1987; Kohn, Paulus, & Choi, 2011; Paulus & Nijstad, 2003). Valuable evidence exists to guide the practice of group creativity including the effects of incentives and motivation (Shepperd, 1993), the role of cognitive and cultural diversity in creative teams (Shin, Kim, Lee, & Bian, 2012; Stahl, Maznevski, Voigt, & Jonsen, 2010), and the links between creative teams and social capital (Han, Han, & Brass, 2014), to name a few. Nonetheless, many issues remain open about group creativity, often with findings that are inconclusive or contradictory, correlations with no meaningful theoretical explanations, a general lack of multi-level connections for example between cognitive, group and social processes, virtually no connections across disciplinary traditions, and a lack of validation of studies in real settings (Sutton & Hargadon, 1996).

Cellular automata and agent-based simulations have been used in the study of creativity and innovation from a variety of research agendas, ranging from illustration and pedagogical, to aids for designing public policy and forecasting future scenarios (Watts & Gilbert, 2014). Axelrod presented a model of cultural transmission (1997) that has been used subsequently to model group creativity and innovation (Sosa & Gero, 2005; Watts & Gilbert, 2014). Other agent models have been proposed for the study of systemic creativity (Kahl & Hansen, 2015), creative cities (Malik, Crooks, Root, & Swartz, 2015), the effects of peer-review in innovation (Sobkowicz, 2015), creative leadership (Leijnen & Gabora, 2010), the impact of intellectual property on innovation (Haydari & Smead, 2015), and idea generation in multi-level neural models (Iyer et al., 2009).

A number of authors working with computational creativity have articulated the need to include social evaluation in an otherwise solipsistic tradition of modelling individual creativity (Grace, Maher, Fisher, & Brady, 2014; Jordanous, 2012; Sosa & Gero, 2015) Key principles across five theoretical domains have been linked to formulate a framework for the study of systemic creativity that includes social evaluation (Sosa, Gero, & Jennings, 2009), and a functional framework was proposed to support multi-level modelling (Sosa & Gero, 2015). This paper adopts a multi-dimensional approach to examine idea generation of groups, focusing on the process by which team members build on the idea of others (Kohn et al., 2011).

The following section presents an analysis of two prominent studies of creativity from the quantitative and qualitative research cultures where similarities and differences are highlighted, and a common issue of interest is extracted to guide the development of a multi-agent simulation of group creativity.

Building on the ideas of others
Two related studies of group creativity from different research traditions are analysed here. An experimental study of idea generation in teams (Girotra, Terwiesch, & Ulrich, 2010) represents the quantitative culture, whilst a four-year case study of creative collaboration between product designers represents the qualitative culture (Elsbach & Flynn, 2013). Table 2 summarises the main similarities, differences, and identifies a common theme in these studies. The similarities include research questions aimed at collaborative work by creative designers in the context of organisational structures, and a particular focus on examining the sharing and building on ideas. The differences range from the research methods applied, the selection of participants and the design task conditions, the sources and type of data, and how the analysis is carried. Whilst these studies are incommensurable and follow very different criteria of scholarship (Mahoney & Goertz, 2006), both identify a common theme related to the collaborative behaviour of building on the ideas of others, which we study in detail in this paper aided by computational simulations described below. This issue is tackled in different ways between research cultures, in the quantitative camp a rating system is applied by trained judges to measure *build-up* scores, whilst in the qualitative tradition, a theory-based framework is built from the data in an iterative process validated by the own designers participating in the study.

*Table 2. Analysis of two studies of group creativity from different research cultures*

|  | (Girotra et al., 2010) | (Elsbach & Flynn, 2013) |
|---|---|---|
| Key similarities | Research question: "How might individual work in combination with working together (hybrid groups) offer advantages over a pure team structure?"<br><br>Motivation: "We build theory that relates organizational phenomena to four different variables that govern the underlying statistical process of idea generation and selection."<br><br>Hypothesis: "The quality of the best ideas generated and selected by a hybrid group is higher than that of a team."<br><br>Results: "We show that idea generation in teams is more likely to lead to ideas that build on each other."<br><br>Limitations: "Our subjects' limited time, resources, and prior exposure to the problem-solving context limit our ability to perfectly mimic a real situation. Furthermore, our subjects were not placed in teams that had developed a deep working relationship." | Research question: "How do specific collaboration activities relate to the self-concepts of creative workers in corporate contexts?"<br><br>Motivation: "We hope to shed light on the reasons why some creative collaborations in organizations succeed, while others fail."<br><br>Hypothesis: "We propose that affirming self-concepts may conflict with collaborative behaviours expected of creative workers."<br><br>Results: "Our analysis indicated that idea-giving behaviours affirmed the personal identities of most designers. The second pattern indicated that idea-taking behaviours threatened personal identities of most designers."<br><br>Limitations: "Further research is needed to identify the extent to which our findings can generalize across different organizations and different organizational cultures." |
| Key differences | Research method: "Laboratory experiment which compares group structures with respect to each of the four variables individually, and which measures their collective impact on the quality of the best idea."<br><br>Data collection: "A within-subjects design, each subject generates ideas under both the treatments -team and hybrid."<br><br>Participants: "44 subjects from an upper-level (undergraduate) product design elective course (who) received training in idea generation techniques. Subjects were randomly divided in groups of four."<br><br>Design problem: "Each team is given 30 minutes to complete an idea generation challenge. In the hybrid treatment, 10 minutes to work individually and an additional 20 minutes to share and discuss their ideas and to develop new ideas. Challenges: "1) A manufacturer of sports and fitness products is interested in new product concepts that might be sold to students in a sporting goods retailer. 2) A manufacturer of dorm and apartment products is interested in new product concepts that might be sold to students in a home-products retailer." | Research method: "Qualitative methods, including interviews and non-participant observation."<br><br>Data collection: "Over a four-year period: 40 open-ended interviews and approximately 100 hours of observation."<br><br>Participants: "40 designers (35 men, 5 women; average age = 39.5 years; average time working at the corporation = 12.7 years). Participants held the titles of staff designer (7), project designer (16), or designer (17)."<br><br>Design problem: "Designers in this division were required to come up with approximately 1000 new toy designs a year. These original designs sometimes involved relatively minor modifications to existing designs (a new toy car design), but often involved the creation of completely new toy concepts."<br><br>Evaluation: "These designers viewed themselves, and were viewed by management, as creative workers. The director also told us that collaborative teamwork was the norm on all design projects, and that designers were expected to collaborate with everyone on a project team. All of the designers reported that they were required to work extensively in collaborative teams and that they considered |

|  | Evaluation: "An accurate measurement of idea quality is central to our work (…) we use two approaches: a quality evaluation tool, which collects about 20 ratings per idea, and a purchase-intent survey, which captures about 40 consumer opinions about their intent to purchase a product. A (judging) panel received formal training in the valuation of new products. To verify the reliability of these ratings, we constructed Kappa and AC1 statistics for each of the two idea domains."<br><br>Units of analysis: "Average quality of ideas, number of ideas, variance in the quality of ideas, and the ability of groups to discern the quality of ideas. Our metric for effectiveness is the quality of ideas selected as the *best*."<br><br>Data analysis: "All hypotheses related to idea quality are tested using both business value and purchase intent as measures of quality. We use an analysis of variance (ANOVA) of the judges' ratings given each idea. We include controls for the four-person group of individuals generating the ideas and the rater who provided the rating. This is because there are substantial differences in ability across the groups, and because there are systemic differences in how the scales were used by different raters." | themselves to be creative workers due to their job requirements to produce original toy designs."<br><br>Units of analysis: "Two primary types of personal identities of designers: (1) 'artistic' personal identities, that included the self-categorizations of 'creator', 'controller', and 'idealist'; and (2) 'problem solving' personal identities, that included the self-categorizations of 'pragmatist', 'refiner', and 'enabler'."<br><br>Data analysis: "Data analysis followed an iterative approach, moving back and forth between theory- development, data review, and literature review. We analysed our data in four stages. We asked (the designers) if our placement made sense. All 30 of the designers confirmed our choice of self-categorization dimensions and personal identity categorizations. Further, many of them confirmed our placements of their colleagues. Based on this feedback, we felt very confident in our descriptions of these 30 designers' personal identity categorizations and that our coding scheme would produce accurate results." |
| Common theme: "Building on the ideas of others" | Definition: "One person can build on the ideas of another in a way that increases the quality of the ideas. As far as we know there has not been any theoretical or empirical support for the claim that these ideas are better than ideas that are generated independently. In our study, we directly explore the role of buildup. To do this, we hired three independent judges to code the substance of ideas on three dimensions: the type of product, the principal sporting activity associated with the product, and the key benefit proposition of the proposed product To construct our buildup metric, we compare the classification of two consecutively generated ideas. For example, if the idea shares all three dimensions with the preceding idea, it earns a buildup score of 3. More generally, the buildup score is the number of dimensions on which an idea shares a value with the idea generated immediately previously. We average this buildup score across the three independent judges."<br><br>Findings: "We show that idea generation in teams is more likely to lead to ideas that build on each other (however) we find that ideas that build on a previous idea are worse, not better, on average. We found that differences in performance across individuals are large and highly significant."<br><br>Managerial implications: "If the interactive buildup is not leading to better ideas, an | Definition: "We used past research as a guide to label instances randomly selected from the interview data. We came to agree upon a framework of six common collaborative behaviours described by designers: two related to giving ideas (i.e. offering ideas and promoting ideas), three related to taking ideas (i.e. soliciting ideas, considering ideas, and incorporating ideas), and one involved taking and giving (co-creating) concurrently."<br><br>Findings: "Our findings suggest that the positive impact of the collaborative behaviour of incorporating ideas may depend, in part, on the types of ideas being incorporated."<br><br>Managerial implications: "Promoting the behaviour of idea-taking (not idea-giving) may be what is critical to improving creative collaborations. These findings make clear that a one-size-fits-all approach to designing and promoting creative collaborations is unlikely to work. "Our findings suggest that, rather than rewarding employees for offering 'the most creative idea', organizations might offer rewards to employees who are able to effectively incorporate others' ideas into their own work. Further, training programmes may need to be updated to include education on how to effectively take ideas from others, in addition to how to give ideas. Effectively leading creative collaborations may mean preventing any single group member from 'owning' an idea in the early stages." |

|  | organization might be better off relying on asynchronous idea generation by individuals." |  |

The importance of "idea build-up" is recognised in both studies, albeit they approach it from different angles, and recognise the need for further research on this theme. How to reason about "idea build-up" and how to tackle it in systematic studies? How do 'idea-taking' and 'idea-giving' shape creative collaboration? How may the definition of "idea build-up" help interpret the outcomes of ideation teams? We propose here a mixed-method approach using computational modelling where a variety of lenses can be worn to look at idea build-up.

**An agent-based simulation of creative collaboration**

An influential simulation model of idea transmission is extended here that aims at examining "just how much of cultural emergence and stability can be explained without resorting to centralized authority" (Axelrod, 1997). In that model of cultural convergence (i.e., imitation), agents share information by exchanging values between neighbours. These stochastic systems gradually converge either to a single value (system ergodicity), or they form 'regions' or clusters when agent interaction is conditioned by compatibility (neighbouring agents that share at least one value at initial time). This simple model has been widely extended, including in studies of creativity and innovation (Araujo & Mendes, 2009; Kiesling, Günther, Stummer, & Wakolbinger, 2012; Leydesdorff, 2002; Sosa & Gero, 2005; Watts & Gilbert, 2014).

3.1 The Model

Axelrod's description of "the entire dynamics of the system" consists of two steps[1]: pick a random agent to be active and one of its neighbours, and with probability equal to their similarity (shared values), these agents interact by selecting at random a feature on which the active agent and its neighbour differ (if any), as a result the active agent changes its trait on this feature to the neighbour's trait on this feature (Axelrod, 1997). The model is usually analysed consisting of hundreds of agents arranged in a two-dimensional grid initialised with randomised cultural values, where 'culture' is defined by an array of integer variables (of size = feature length) and the domain of these variables (traits) typically {0, 1, 2… 9}. Notice that in the original formulation of this model, all active agents are defined as idea-taking agents (Elsbach & Flynn, 2013).

The extension to this model presented here consists of incorporating a type of change agency based on dissent or divergent behaviour. Applying the MDC framework of creativity (Sosa & Gero, 2015), dissent is modelled at the societal level (MDC-S) as a standard behaviour where every agent that identifies total group convergence attempts to introduce a new idea

---

[1] "At random, pick a site (cell) to be active, and pick one of its neighbors. With probability equal to their cultural similarity, these two sites interact. An interaction consists of selecting at random a feature on which the active site and its neighbour differ (if there is one) and changing the active site's trait on this feature to the neighbor's trait on this feature" (Axelrod, 1997)

by changing its variables to random values (with a low probability of success of 0.1%). Inspired by the two ideation studies analysed above, we distinguish between idea-giving and idea-taking behaviours at the group level (MDC-G) by defining a ratio from each type in the formation of groups at initial time in the simulation. The specific aim here is to understand the effects of group composition across different ratios of 'idea takers' and 'idea givers' interacting.

Previous studies have shown that incorporating dissent or divergent behaviour in the original Axelrod model triggers cycles of collective convergence -or consensus on a dominant value followed by sudden episodes of collective change where groups adopt new values (Sosa & Gero, 2005). This paper investigates the effects of different ratios of idea-taking and idea-giving agents in a group. To this end, we repeatedly run the model to obtain average values for groups where all agents are 'idea-takers' (as the original formulation), and gradually introduce idea-giving agents until groups where all agents are 'idea-givers'. One specific dependent variable is measured: the number of collective changes of the dominant value, which we call a *revolution* (Kuhn, 1996). In this context, a revolution is registered whenever the entire group of agents adopts a value that was introduced by a dissenting agent. The model analysed here consists of 9 agents in a 3x3 torus grid. Following Axelrod's lexicon, ideas are encoded in 6 features, with 9 traits, with 'Von Neumann' neighbourhoods, and initialised in fully converged mode, i.e., all agents adopting the same ideas at initial state. Results are obtained by averaging $10^3$ cases running for $10^4$ simulation steps.

## Results

Group composition is inspected from "all takers" (9t) to "all givers" (9g) in the group, in increments of one member each time. The effects of group composition are not linear -as shown in Figure 1. First, *revolutions* (change episodes) are highest in groups where all active agents are idea-takers, or "rev_9t" in Figure 1 (as in Axelrod's original model). Introducing idea-giving agents in the group affects the indicator of group creativity in this model by first causing a *dip* in revolutions when one idea-giver agent is introduced ("rev_8t1g" in Figure 1), then increasing as a majority of agents engage in idea-giving behaviour.

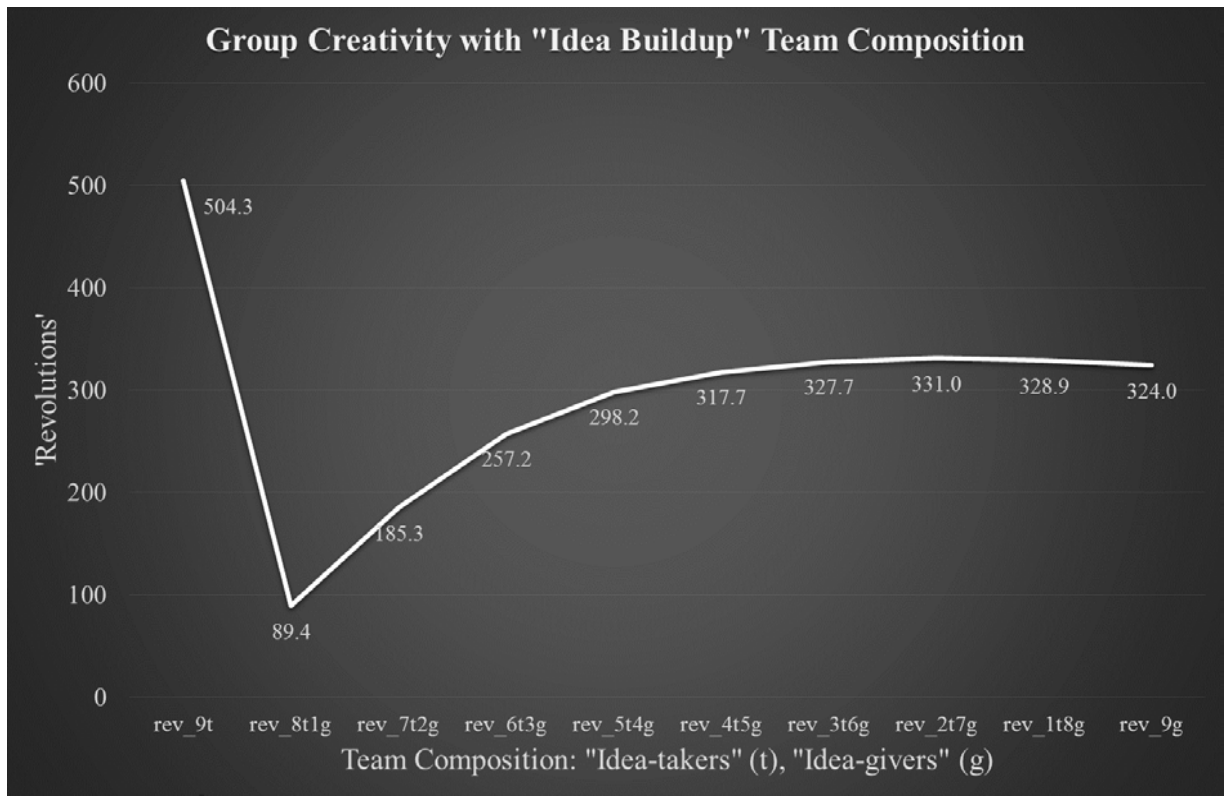

*Figure 1. Quantitative analysis of the idea-taking and idea-giving model (averages)*

A way to interpret these results is to consider that a "9t" scenario is only a theoretical construct, as research shows that all members of creative design teams engage in idea-giving (Elsbach & Flynn, 2013). It is also worth noting that the increase in group creativity also creates a different *type* of revolutions: when new ideas gain dominance, it is possible to analyse whether they are adopted as introduced by the change agent, or whether they are the product of combinatorial contributions from multiple team members. When only a few idea-giving agents are present in the team and the level of revolutions is low, change agents tend to retain 'ownership' of new ideas, whereas in groups with more idea-giving agents where revolutions take place more often, winning ideas are almost by definition result of 'idea build-up'. This seems like a fertile ground for co-creation, but it could also lead to negative effects such as *groupthink* (Rose, 2011) in the absence of appropriate facilitation strategies. These quantitative results give a general overview of the effects of idea-taking and idea-giving when many cases are averaged. However, it is far from clear from looking at these results what may explain the significant *dip* in group creativity with the introduction of one idea-giving team member. Why are (hypothetical) teams of all idea-takers able to trigger collective change more easily than teams where only one idea-giver agent is inserted? And why is this result *reversed* as more idea-givers are introduced in a team?

To examine these questions, we adopt a qualitative lens to look at a specific case across three conditions: when all team members engage in idea-taking ("rev_9t"), when only one idea-giver joins the team ("rev_8t1g"), and when three idea-giver agents interact ("rev_6t3g"). Figure 2 shows a specific revolution episode from each of these three conditions that illustrates the *type* of revolutions that occur. All episodes start with a new idea being introduced breaking group convergence: each frame shows a simulation step, nine agents in a torus 3x3 grid, each agent has 3 features and 9 possible traits per feature. In rev_9t, all agents have adopted idea [000] when agent #4 introduces [626]. In rev_8t1g, all agents hold [000] when agent #7 introduces [238], and in rev_6t3g, agent #2 introduces [682]. In rev_9t, the second feature of the new idea spreads through the group, reaching dominance as [020] in a few steps. In rev_8t1g, the new idea starts spreading as three composite ideas [200], [030], and [230], and all agents switch to one of these variants, except for agent #0, which sticks to [000], i.e., this is the agent fixed on idea-giving behaviour. In rev_6t3g, the new idea [682] spreads as [600], [002], and [602], ultimately gaining dominance as [002] in the group.

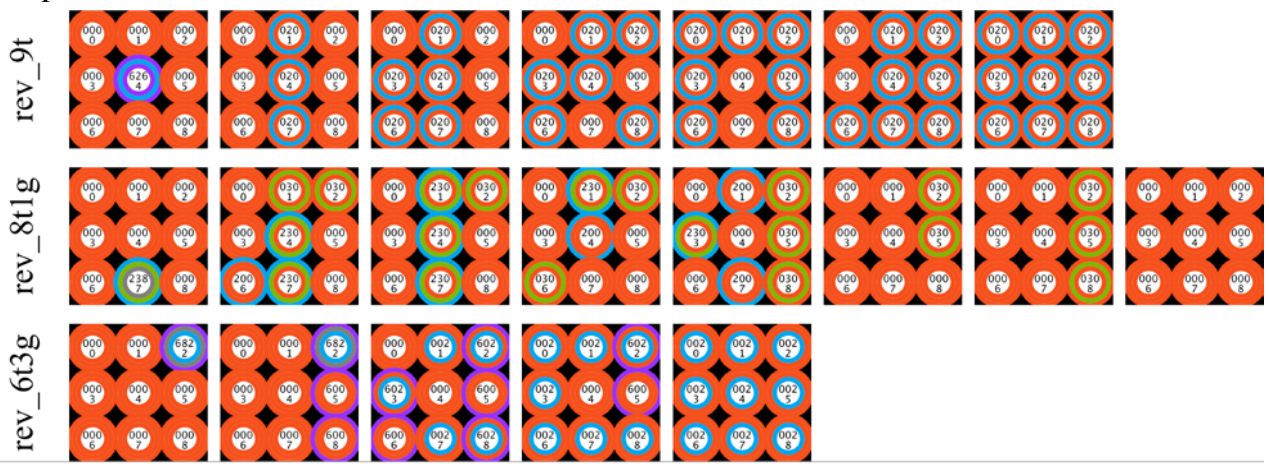

*Figure 2. Qualitative analysis of the idea-taking and idea-giving model (one selected case)*

These results are suggestive of the complexities of group creativity: when all agents in a team engage in idea-taking, the probability that a new value be spread is high because at any given iteration step, when agents interact with their neighbours, active agents *take* the new 'dissenting' value from neighbours. However, when an idea-giving agent is introduced, it controls the process and becomes a 'gatekeeper' by virtue of being the only agent in the group to give ideas to others. Such *monopoly* is broken as soon as another team member has access to idea-giving capabilities, as both of them take ideas from others when they are in a neighbouring (inactive) position of the other 'idea-giving' (active) agent. In teams of 'all givers', the 'taking' role is performed by all agents when in the role of neighbours of active agents.

Thus we can see that, other than the original Axelrod conditions of fully imitative team members, this model shows *that* teams with only a single idea-giving member support very low levels of group creativity. The model also shows that a critical mass of idea-giving behaviour may be sufficient to reach high levels of group creativity, after which more idea-giving fails to increase group creativity beyond what appears to be a 'glass ceiling' or natural group capacity. Moreover, the model shows *why* this occurs: teams with a single dominant idea-giving role suffer from a bottleneck enforced by an agent that indirectly adopts a 'change resistance' role. Rather than removing that type of agent from a team, our model suggests that adding another idea-giving agent helps to balance the interactions and rapidly increases the creative potential of the group. After a critical mass of about half the team, adding more idea-giving agents may not have an impact on group creativity, and it may only be beneficial if an increase in co-creation strategies is preferred.

Although these results cannot be generalised beyond this model, it is a good illustration of the sensitivity of creative collaboration, and the large effects that even apparently small changes can have. There are caveats to these results, with fundamental managerial implications: whilst the *dip* between "all idea-takers" and "one idea-giver" is significant, only 58% of cases show a decrease in number of revolutions, whilst 29% remain constant and 13% in fact show an increase in this indicator of group creativity. This should be considered a strength rather than a weakness of this simple model, as its stochastic nature is reminiscent of the variability and unpredictable nature of creativity across cases.

## Discussion

We have applied agent-based simulations as 'intuition pumps' to think quantitatively and qualitatively about group creativity, and in particular 'idea build-up', or the building of new ideas on the ideas of others. Big claims about the outcomes of this model in relation to real world teams are explicitly avoided, as their role here is only to guide our intuitions. The main lessons to think about creative groups are: a) group formation can be critical when initial conditions define ranges of possible outcomes, and effects can be expected to be non-linear; b) the balance between idea-giving and idea-taking is a delicate one in creative collaborations, and a range of consequences should be considered both in research and practice, such as idea ownership and idea decomposition; c) finally, this work shows that agent-based models are useful as intuition pumps to reason about complex situations such as creativity and innovation.

## References

Araujo, T., & Mendes, R. V. (2009). Innovation and Self-Organization in a Multi-Agent Model. *Advances in Complex Systems*, *12*(02), 233–253. http://doi.org/10.1142/S0219525909002180
Axelrod, R. (1997). The Dissemination of Culture: A Model with Local Convergence and Global Polarization. *Journal of Conflict Resolution*. http://doi.org/10.1177/0022002797041002001
Cardoso de Sousa, F. (2007). Still the elusive definition of creativity. *International Journal of Psychology: A Bio Psychosocial*, *2*, 55–82.
Creswell, J. W. (2009). *Research design: Qualitative, quantitative, and mixed methods approaches*. *Research Design qualitative quantitative and mixed methods approaches* (Vol. 3rd). http://doi.org/10.1016/j.math.2010.09.003
Dennett, D. C. (2014). *Intuition Pumps And Other Tools for Thinking*. W. W. Norton & Company. Retrieved from https://books.google.com/books?hl=en&lr=&id=9SduAwAAQBAJ&pgis=1
Diehl, M., & Stroebe, W. (1987). Productivity loss in brainstorming groups: Toward the solution of a riddle. *Journal of Personality and Social Psychology*. http://doi.org/10.1037/0022-3514.53.3.497
Elsbach, K. D., & Flynn, F. J. (2013). Creative Collaboration and the Self-Concept: A Study of Toy Designers. *Journal of Management Studies*, *50*(4), 515–544. http://doi.org/10.1111/joms.12024
Francis, D., & Bessant, J. (2005). Targeting innovation and implications for capability development. *Technovation*, *25*(3), 171–183. http://doi.org/10.1016/j.technovation.2004.03.004
Fruchter, R. (2001). Dimensions of teamwork education. *International Journal of Engineering Education*, *17*(4 and 5), 426–430.

Girotra, K., Terwiesch, C., & Ulrich, K. T. (2010). Idea Generation and the Quality of the Best Idea. *Management Science*, *56*(4), 591–605. http://doi.org/10.1287/mnsc.1090.1144

Grace, K., Maher, M. L., Fisher, D., & Brady, K. (2014). Modeling expectation for evaluating surprise in design creativity. In *Design Computing and Cognition'14* (pp. 189–206). Springer International Publishing.

Han, J., Han, J., & Brass, D. J. (2014). Human capital diversity in the creation of social capital for team creativity. *Journal of Organizational Behavior*, *35*(1), 54–71. http://doi.org/10.1002/job.1853

Haydari, S., & Smead, R. (2015). Does Longer Copyright Protection Help or Hurt Scientific Knowledge Creation? *Journal of Artificial Societies and Social Simulation2*, *18*(2), 23.

Iyer, L. R., Doboli, S., Minai, A. A., Brown, V. R., Levine, D. S., & Paulus, P. B. (2009). Neural dynamics of idea generation and the effects of priming. *Neural Networks : The Official Journal of the International Neural Network Society*, *22*(5-6), 674–86. http://doi.org/10.1016/j.neunet.2009.06.019

Jordanous, A. (2012). A Standardised Procedure for Evaluating Creative Systems: Computational Creativity Evaluation Based on What it is to be Creative. *Cognitive Computation*, *4*(3), 246–279. http://doi.org/10.1007/s12559-012-9156-1

Kahl, C. H., & Hansen, H. (2015). Simulating Creativity from a Systems Perspective: CRESY. *Journal of Artificial Societies and Social Simulation*, *18*(1), 4. Retrieved from http://jasss.soc.surrey.ac.uk/18/1/4.html

Kiesling, E., Günther, M., Stummer, C., & Wakolbinger, L. M. (2012). Agent-based simulation of innovation diffusion: A review. *Central European Journal of Operations Research*. http://doi.org/10.1007/s10100-011-0210-y

Kohn, N. W., Paulus, P. B., & Choi, Y. (2011). Building on the ideas of others: An examination of the idea combination process. *Journal of Experimental Social Psychology*, *47*(3), 554–561. http://doi.org/10.1016/j.jesp.2011.01.004

Kuhn, T. S. (1996). The structure of scientific revolutions. *The University of Chicago Press*, *3rd Ed.* Retrieved from papers2://publication/uuid/C5D275F2-36F4-44E2-BA6E-BCEF8D99CE3B

Leijnen, S., & Gabora, L. (2010). An agent-based simulation of the effectiveness of creative leadership. In *Proceedings of the Annual Meeting of the Cognitive Science Society*. Portland, Oregon.

Leydesdorff, L. (2002). The complex dynamics of technological innovation: a comparison of models using cellular automata. *Systems Research and Behavioral Science*, *19*(6), 563–575.

Mahoney, J., & Goertz, G. (2006). A tale of two cultures: Contrasting quantitative and qualitative research. *Political Analysis*, *14*(3), 227–249. http://doi.org/10.1093/pan/mpj017

Malik, A., Crooks, A., Root, H., & Swartz, M. (2015). Exploring Creativity and Urban Development with Agent-Based Modeling. *Journal of Artificial Societies and Social Simulation*, *18*(2), 12.

Osborn, A. F. (1963). *Applied Imagination: Principles and procedures of creative problem solving*. Oxford. http://doi.org/citeulike-article-id:975273

Paulus, P. B., & Nijstad, B. A. (2003). *Group Creativity: Innovation through Collaboration. Work* (Vol. 12). http://doi.org/10.1016/S1572-0977(06)12002-6

Rose, J. D. (2011). Diverse Perspectives on the Groupthink Theory – A Literary Review. *Emerging Leadership Journeys*, *4*, 37–57. Retrieved from http://www.regent.edu/acad/global/publications/elj/vol4iss1/Rose_V4I1_pp37-57.pdf

Shepperd, J. A. (1993). Productivity loss in performance groups: A motivation analysis. *Psychological Bulletin*. http://doi.org/10.1037/0033-2909.113.1.67

Shin, S. J., Kim, T. Y., Lee, J. Y., & Bian, L. (2012). Cognitive team diversity and individual team member creativity: A cross-level interaction. *Academy of Management Journal*, *55*(1), 197–212. http://doi.org/10.5465/amj.2010.0270

Sobkowicz, P. (2015). Innovation Suppression and Clique Evolution in Peer-Review-Based, Competitive Research Funding Systems: An Agent-Based Model. *Journal of Artificial Societies and Social Simulation2*, *18*(2), 13.

Sosa, R., & Gero, J. S. (2005). A Computational Study of Creativity in Design: The Role of Society. *Artificial Intelligence for Engineering Design, Analysis and ManufacturingAI EDAM*. http://doi.org/10.1017/S089006040505016X

Sosa, R., & Gero, J. S. (2015). Multi-dimensional creativity: a computational perspective. *International Journal of Design Creativity and Innovation*, 1–25. http://doi.org/10.1080/21650349.2015.1026941

Sosa, R., Gero, J. S., & Jennings, K. (2009). Growing and destroying the worth of ideas. In *Proceeding of the seventh ACM conference on Creativity and cognition - C&C '09* (p. 295). New York, New York, USA: ACM Press. http://doi.org/10.1145/1640233.1640278

Stahl, G. K., Maznevski, M. L., Voigt, A., & Jonsen, K. (2010). Unraveling the effects of cultural diversity in teams: A meta-analysis of research on multicultural work groups. *Journal of International Business Studies*. http://doi.org/10.1057/jibs.2009.85

Sutton, R., & Hargadon, A. (1996). Brainstorming Groups in Context : Effectiveness in a Product Design Firm. *Administrative Science Quarterly*, *41*(4), 685–718. http://doi.org/10.2307/2393872

Watts, C., & Gilbert, N. (2014). *Simulating Innovation: Computer-based Tools for Rethinking Innovation*. Edward Elgar Publishing.

**Author Biographies**

Ricardo Sosa

Ricardo combines a creative background as a designer with a passion for the study of computational systems. He studies creativity and innovation principles through multi-agent social systems and is involved in the development of facilitation practices for team ideation and for participatory decision making. Ricardo partners with colleagues across disciplines including: robotics, social science, cognitive science, architecture, arts, engineering, business, public health, and computer science: https://colab.aut.ac.nz/staff/ricardo-sosa

Andy Connor

Andy is a mechanical engineer by training but has a breadth of experience in mechatronics, software engineering, computer science and more recently in creative technologies. Andy has a broad range of research interests that include automated design, computational creativity, education, evolutionary computation, machine learning and software engineering: https://www.aut.ac.nz/profiles/creative-technologies/senior-lecturers/andy-connor